# Local Moment Formation by Vacancies in Mono-layer Graphene

Partha Goswami and Ajay Pratap Singh Gahlot

*Deshbandhu College, University of Delhi, Kalkaji, New Delhi-110019,India*

physicsgoswami@gmail.com;Tel:0091-129-243-9099.

**Abstract** We employ the Green's function technique to investigate the vacancy-induced quasi-localized magnetic moment formation in monolayer graphene starting with the Dirac Hamiltonian, which focuses on the π-orbitals only, involving the nearest neighbor(NN)(t) and moderate second neighbor(SN)(t′ < t/3) hopping integrals. The vacancy defect is modeled by the addition of the on-site perturbation potential to the Hamiltonian. We find that, when (t′/t) << 1, the vacancy induced π-state at the zero of energy(zero-mode state(ZMS)) does not inhabit the minority sub-lattice due to the strong scalar potential induced by the vacancy(the ZMSs get lodged in the majority sub-lattice) whereas, when (t′/t) is increased, the ZMS is somewhat suppressed. This shows that, not only the shift of the Fermi energy away from the linearly-dispersive Dirac points, the issue of this topological localization is also hinged on the ratio (t′/t). Furthermore, when a vacancy is present, the three $sp^2$-hybridized σ states of each of the three nearest-neighbor carbon atoms, forming a carbon triangle surrounding the vacancy, are close to the Fermi energy ($E_F$). The Hund's coupling between these σ electrons and the remaining electron which occupies the π state spin polarizes the π state leading to local moment formation close to $E_F$. Since the system at the Fermi level has low electronic density, there is poor screening of such magnetic moments. This may lead to a high Curie temperature for such vacancy-induced moments.

## 1. Introduction

The monolayer graphene is a plane of $sp^2$ hybridized carbon atoms arranged in a honeycomb lattice (see Figure A) which makes the system a half-filled one with a density of states that vanishes at the neutrality point. The effective, low energy quasi-particle spectrum of the system is characterized by a dispersion which is linear in momentum [1] close to the Fermi energy. The latter implies that the corresponding quasi-particles would behave as Dirac massless chiral fermions. As a result of which many exotic phenomena, such as the Klein tunneling[2], the zitterbewegung in confined structures[3], etc. in the domain of quantum electrodynamics find a practical realization in this solid state material. The review of the consequence of ever-present vacancies in graphene owing to its exposed surface and the substrates is the focal point of the present communication. The posting is, in fact, an exercise leaning very heavily on refs. [4],[5] and [6]. It is also inspired by the recent reporting of Ugeda et al. [7] where atomic vacancies have been artificially generated on a graphite surface and formation of quasi-localized magnetic moments corresponding to the state of zero-energy with considerable reduction in the carrier mobility has been observed in the neighborhood of these vacant sites.

The electronic structure of mono-layer graphene can be captured within a tight-binding approach, in which the Dirac electrons are allowed to hop between the nearest neighbors (NN) with hopping integral t = 2.8 eV, and also between next nearest neighbors(2NN) with an additional hopping t′ (t′ < t/3). The inclusion of the second term introduces an asymmetry between the valence and conduction bands leading to the violation of the particle-hole symmetry. We have reviewed the formation of vacancy-induced quasi-localized magnetic moment formation in this system starting with the Hamiltonian involving these hopping integrals and using the finite-temperature Green's function formalism (see section 2). The vacancy defect is modeled by the addition of the on-site perturbation potential to the Hamiltonian (see section 3). As regards the wave function of the impurity state, it is shown in section 4 that if practically the NN hopping is present in a bipartite lattice, a single vacancy in one sub-lattice (this may be referred to as the minority sub-lattice since there is one atom less) introduces a state at the zero of energy and this state does not live on the minority sub-lattice (The state, in fact, is lodged on the majority sub-lattice as shown by Castro Neto et al. [4] using the rank-nullity theorem in linear algebra.). The first result follows from the approximate particle-hole symmetry whereas the second result follows by examining the perturbed wave function in the presence of the impurity potential($U_0$) analyzing the Lippmann-Schwinger equation. The equation relates the perturbed wave function to the unperturbed wave function via the Green's function $G^0$ : $|\psi\rangle = |\psi^0\rangle + G^0 U_0 |\psi\rangle$. We obtain

$$|\psi\rangle = |\psi^0\rangle + G^0 U_0 |\psi^0\rangle + G^0 U_0 G^0 U_0 |\psi\rangle = |\psi^0\rangle + \{G^0 U_0 |\psi^0\rangle/(1 - G^0 U_0)\}.$$

From this equation, one may also estimate the spread of the wave function to the minority sub-lattice if the 2NN interactions are present. We next provide in section 4 a heuristic but compelling argument (see also refs.5 and 6) to clarify how the magnetic moment formation takes place with an isolated vacancy. The argument is based on the fact that the $sp^2$ – hybridized σ states of a carbon atom (in graphene), ordinarily omitted from the tight- binding graphene Hamiltonians (which focus on the π orbitals), are very important when a vacancy is present as these states get introduced near the Fermi energy. They control the electronic behavior of the vacancy as explained in brief in section 4.

## 2. Unperturbed propagators of mass-less chiral fermions

The Dirac mass-less chiral fermions in mono-layer graphene are described by the Hamiltonian

$$H_0 = \sum_{\mathbf{k}} (a^\dagger_{\mathbf{k}} \quad b^\dagger_{\mathbf{k}}) \begin{bmatrix} \beta(k) & \gamma^*(k) \\ \gamma(k) & \beta(k) \end{bmatrix} \begin{pmatrix} a_k \\ b_k \end{pmatrix} \qquad (1)$$

where $\gamma = \gamma(k) = -t\sum_{j=1,2,3}\exp(ik.\delta_j)$. The operators $a^\dagger_k$ and $b^\dagger_k$, respectively, correspond to the fermion creation operators for A and B sub-lattices in the mono-layer. The valley and the spin degeneracy factors ($g_v = g_s = 2$) are to be included later on in the density of states. The $\delta_j$'s are three nearest neighbor vectors, $\delta_1 = (a/2)(\mathbf{1}, \sqrt{3})$, $\delta_2 = (a/2)(\mathbf{1}, -\sqrt{3})$, and $\delta_3 = a(-\mathbf{1}, \mathbf{0})$. This leads to $\gamma = \gamma(k) = -t[2 \exp(ik_x a/2) \cos(\sqrt{3}k_y a/2)+\exp(-ik_x a)]$. With the inclusion of the second neighbor hopping (hopping parameter $t' \approx 0.15$ eV), the element $\beta = \beta(\mathbf{k}) = t'\sum_{j=1,2,...6}\exp(ik.d_j) = t'[2\cos(\sqrt{3} k_y a) + 4\cos(\sqrt{3} k_y a/2)\cos(3 k_x a/2)]$ and $d_j$s are the second NN positions given by $d_{1,2} = \pm a_1$, $d_{3,4} = \pm a_2$, $d_{5,6} = \pm(a_2 - a_1)$ where $a_1 = (a/2)(3, \sqrt{3})$ and $a_2 = (a/2)(3, -\sqrt{3})$. One obtains the particle-hole asymmetric energy bands from (1): $E_\pm(k) = \pm|t|[3 + 2\cos(\sqrt{3} k_y a) + 4\cos(\sqrt{3} k_y a/2)\cos(3 k_x a/2)]^{1/2} + t'[2\cos(\sqrt{3} k_y a) + 4\cos(\sqrt{3} k_y a/2)\cos(3 k_x a/2)]$. If $t' < 0$, then $E_\pm(k) = \pm|t|[3 + 2\cos(\sqrt{3} k_y a) + 4\cos(\sqrt{3} k_y a/2)\cos(3 k_x a/2)]^{1/2} - |t'|[2\cos(\sqrt{3} k_y a) + 4\cos(\sqrt{3} k_y a/2)\cos(3 k_x a/2)]$. In what follows we take $t' > 0$.

In order to ultimately describe the formation of vacancy-induced local magnetic moment state formation in the system under consideration, we introduce few unperturbed thermal averages in this section determined by the Hamiltonian in (1), viz. $G^0_{AA}(\mathbf{k},\tau) = -\langle T\{a_\mathbf{k}(\tau) a^\dagger_\mathbf{k}(0)\}\rangle$, $G^0_{AB}(\mathbf{k},\tau) = -\langle T\{a_\mathbf{k}(\tau) b^\dagger_\mathbf{k}(0)\}\rangle$, $G^0_{BA}(\mathbf{k},\tau) = -\langle T\{b_\mathbf{k}(\tau) a^\dagger_\mathbf{k}(0)\}\rangle$, and $G^0_{BB}(\mathbf{k},\tau) = -\langle T\{b_\mathbf{k}(\tau) b^\dagger_\mathbf{k}(0)\}\rangle$ as a preliminary measure. The vacancy defect, modeled by the addition of the on-site perturbation potential to the Hamiltonian, will lead to the perturbed propagators as will be shown in the next section. Here T is the time-ordering operator which arranges other operators from right to left in the ascending order of imaginary time $\tau$. The first step of the scheme of calculation of real-space unperturbed propagators involves the consideration of (imaginary) time evolution of the operators $a_\mathbf{k}(\tau)$ where, in units such that $\hbar = 1$, $a_\mathbf{k}(\tau) = \exp(H\tau)a_\mathbf{k}(0)\exp(-H\tau)$. We obtain, for example, $\partial a_\mathbf{k} = -\beta a_\mathbf{k} - \gamma^* b_\mathbf{k}$, $\partial b_\mathbf{k} = -\beta b_\mathbf{k} - \gamma a_\mathbf{k}$, and so on. Here $\partial \equiv (\partial/\partial\tau)$ and the argument part has been dropped in writing the operators ($a_\mathbf{k}(\tau), b_\mathbf{k}(\tau)$) and their derivative. As the next step, we find that the equations of motion of the averages are given by $\partial G^0_{AA}(\mathbf{k},\tau) = -\beta G^0_{AA}(\mathbf{k},\tau) - \gamma^* G^0_{BA}(\mathbf{k},\tau) - \delta(\tau)$, $\partial G^0_{BA}(\mathbf{k},\tau) = -\beta G^0_{BA}(\mathbf{k},\tau) - \gamma G^0_{AA}(\mathbf{k},\tau)$, $\partial G^0_{BB}(\mathbf{k},\tau) = -\beta G^0_{BB}(\mathbf{k},\tau) - \gamma G^0_{AB}(\mathbf{k},\tau)$, and $\partial G^0_{AB}(\mathbf{k},\tau) = -\beta G^0_{AB}(\mathbf{k},\tau) - \gamma^* G^0_{BB}(\mathbf{k},\tau)$. The third step is the calculation of the Fourier coefficients $G_{\alpha\beta}(\mathbf{k},\omega_n) = \int_0^\beta d\tau\, e^{i\omega_n\tau} G_{\alpha\beta}(\mathbf{k},\tau)$ (where the Matsubara frequencies are $\omega_n = [(2n+1)\pi/\beta]$ with $n = 0, \pm 1, \pm 2,\ldots$) of these temperature Green's functions. Here $\beta = (k_B T)^{-1}$. We refrain from writing explicitly the equations to determine these coefficients as this is a trivial exercise. Upon solving the equations, in view of the Hamiltonian in (1), we obtain (replacing $i\omega_n = (\omega + i\eta)$ where $\eta$ stands for $0^+$)

$$G^0_{AA}(\mathbf{k},\omega) = G^0_{BB}(\mathbf{k},\omega) = (\omega + i\eta - E_+(\mathbf{k}))^{-1} (\omega + i\eta - E_-(\mathbf{k}))^{-1}(\omega + i\eta - E_+(\mathbf{k}) + |\gamma(\mathbf{k})|),$$

$$= (1/2)[(\omega + i\eta - E_+(\mathbf{k}))^{-1} + (\omega + i\eta - E_-(\mathbf{k}))^{-1}],$$

$$G^0_{AB}(\mathbf{k},\omega) = \gamma*(\mathbf{k}) (\omega + i\eta - E_+(\mathbf{k}))^{-1} (\omega + i\eta - E_-(\mathbf{k}))^{-1}$$

$$= (\gamma*(\mathbf{k})/2 |\gamma(\mathbf{k})|)[(\omega + i\eta - E_+(\mathbf{k}))^{-1} - (\omega + i\eta - E_-(\mathbf{k}))^{-1}],$$

$$G^0_{BA}(\mathbf{k},\omega) = \gamma(\mathbf{k}) (\omega + i\eta - E_+(\mathbf{k}))^{-1} (\omega + i\eta - E_-(\mathbf{k}))^{-1}$$

$$= (\gamma(\mathbf{k})/2 |\gamma(\mathbf{k})|)[(\omega + i\eta - E_+(\mathbf{k}))^{-1} - (\omega + i\eta - E_-(\mathbf{k}))^{-1}]. \quad (2)$$

The final step is the calculation of the real space propagators. For example, the propagator $G^0_{0,A;\,0,A}(\omega)$ which refers to the central cell(denoted by the index '0') and the sub-lattice 'A' is given by $G^0_{0,A;\,0,A}(\omega) = \Omega_{BZ}^{-1} \int d\mathbf{k}\, \exp[i\mathbf{k}\cdot(\mathbf{r}_{0,A} - \mathbf{r}_{0,A})]\, G^0_{AA}(\mathbf{k},\omega) = \Omega_{BZ}^{-1} \int d\mathbf{k}\, (1/2)[(\omega + i\eta - E_+(\mathbf{k}))^{-1} + (\omega + i\eta - E_-(\mathbf{k}))^{-1}] = F_0(\omega) - i\pi \rho^{(0)}_{0A}(\omega)$, where $F_0(\omega)$ is the real part and $(-\pi^{-1} \rho^{(0)}_{0A}(\omega))$ is the imaginary part of the propagator. Using the result $(x \pm i\eta)^{-1} = [P(x^{-1}) \pm (1/i)\pi \delta(x)]$, where P represents a Cauchy's principal value, one may write

$$G^0_{0,A;\,0,A}(\omega) = \Omega_{BZ}^{-1} \int d\mathbf{k}\, (1/2)[\, P(\omega - E_+(\mathbf{k}))^{-1} - i\pi \delta(\omega - E_+(\mathbf{k}))$$

$$+ P(\omega - E_-(\mathbf{k}))^{-1} - i\pi \delta(\omega - E_-(\mathbf{k}))]. \quad (3)$$

This enables us to write $\rho^{(0)}_{0A}(\omega)$ as the sum of two $\delta$-functions: $\rho^{(0)}_{0A}(\omega) = \Omega_{BZ}^{-1} \int d\mathbf{k}\, \rho_0(\mathbf{k},\omega)$, $\rho_0(\mathbf{k},\omega) = (1/2)[\delta(\omega - E_+(\mathbf{k})) + \delta(\omega - E_-(\mathbf{k}))]$. Now upon taking the real part $F_0(\omega) = \Omega_{BZ}^{-1} \int d\mathbf{k}\, (1/2)\, P[(\omega - E_+(\mathbf{k}))^{-1} + (\omega - E_-(\mathbf{k}))^{-1}]$ and using the expression of $\rho_0(\mathbf{k},\omega)$, in terms of $\delta$-functions, here, one may write

$$F_0(\omega) = (1/2)\, \Omega_{BZ}^{-1} \int_{-\infty}^{+\infty} d\varepsilon \int d\mathbf{k}\, \rho_0(\mathbf{k},\varepsilon)\, P[(\omega - E_+(\mathbf{k}))^{-1} + (\omega - E_-(\mathbf{k}))^{-1}]$$

$$= (1/4)\, \Omega_{BZ}^{-1} \int_{-\infty}^{+\infty} d\varepsilon \int d\mathbf{k}\, [\delta(\varepsilon - E_+(\mathbf{k})) + \delta(\varepsilon - E_-(\mathbf{k}))]$$
$$\times P[(\omega - E_+(\mathbf{k}))^{-1} + (\omega - E_-(\mathbf{k}))^{-1}]$$

$$= (1/2) \int_{-\infty}^{+\infty} d\varepsilon\, \rho^{(0)}_{0A}(\varepsilon)\, P[(\omega - \varepsilon)^{-1}] \quad (4)$$

where the formal definition of the Cauchy principal value operator is

$$P\,{}_0\!\!\int^{+\infty} dx\, f(x)\, [(\omega^2 - x^2)^{-1}] = \lim_{\eta \to 0} [{}_0\!\!\int^{\omega-\eta} dx + {}_{\omega+\eta}\!\!\int^{\infty} dx\,]\, f(x)\,(\omega^2 - x^2)^{-1}.$$

One may note that $\rho^{(0)}_{0A}(\varepsilon)$ = constant is inadmissible as in that case $F_0(\omega)$ = constant $\times \omega\,{}_0\!\!\int^{+\infty} d\varepsilon\, P[(\omega^2 - \varepsilon^2)^{-1}]$. The Cauchy principal value identity is ${}_0\!\!\int^{+\infty} d\varepsilon\, P[(\omega^2 - \varepsilon^2)^{-1}] = 0$. Since the real part equal to zero does not make sense, so indeed $\rho^{(0)}_{0A}(\varepsilon) \neq$ constant.

We choose to write $\rho_0(\mathbf{k},\omega)$ introduced above in terms of Lorentzians. We obtain the unperturbed dimensionless local density of states (LDOS) at the central cell (and the sub-lattice 'A') as $|t|\,\rho^{(0)}_{0A}(\varepsilon/|t|) = \Omega_{BZ}^{-1} \int d(\mathbf{k}a)\, \rho_0(\mathbf{k},\omega)$ where

$$\rho_0(\mathbf{k},\omega) = (\acute{\Gamma}/|t|)[\{1/((\,(\varepsilon/|t|) - E(\mathbf{k}))^2 + (\acute{\Gamma}/|t|)^2)\} + \{1/((\,(\varepsilon/|t|) + E(\mathbf{k}))^2 + (\acute{\Gamma}/|t|)^2)\}],$$

$$E(\mathbf{k}) = [3 + 2\cos(\sqrt{3}\, k_y a) + 4\cos(\sqrt{3}\, k_y a/2)\cos(3\, k_x a/2)]^{1/2}$$

$$+ (t'/|t|)[2\cos(\sqrt{3}\, k_y a) + 4\cos(\sqrt{3}\, k_y a/2)\cos(3\, k_x a/2)].$$

So, in terms of dimensionless quantities, Eq.(4) may be written as

$$[|t|F_0(\omega/|t|)] = (1/2)\,{}_{-\infty}\!\!\int^{+\infty} d(\varepsilon/|t|)\,[|t|\rho^{(0)}_{0A}(\varepsilon/|t|)]\, P[(\omega/|t| - \varepsilon/|t|)^{-1}]. \tag{5}$$

It is crucial for us to calculate $[|t|\rho^{(0)}_{0A}(\varepsilon/|t|)]$. This will yield $F_0(\omega/|t|)$ from Eq.(5). Furthermore, as we shall see in the next section, this will also yield the LDOS at the impurity site calculated using the full Green's function. We have plotted the quantity $[|t|\rho^{(0)}_{0A}(\varepsilon/|t|)]$ as a function of $(\varepsilon/|t|)$ below in Figure 1. For the comparison purpose we have also plotted the square lattice tight band DOS in the Hubbard model which clearly shows van Hove singularity at $(\varepsilon/|t|) = 0$. It may be mentioned that as long as long as $t' < |t|/3$, we find $[|t|\,\rho^{(0)}_{0A}(\varepsilon/|t|)]$ has two peaks at $(\varepsilon/|t|) > 0$ and $(\varepsilon/|t|) < 0$ (see Figure 1(a)). However, when $t'$ is close to $|t|/3$, there is a drastic change in the LDOS (see Figure 1(b)). It resembles the square lattice tight band DOS in the Hubbard model some what. Thus, in monolayer graphene many such issues are hinged on the ratio $(t'/t)$.

**3. The vacancy defect and full Green's function matrix**

As we have noted above, the impurity is modeled here by adding an on-site perturbation V to the Hamiltonian $H = H_0 + V$ where $V = U_0\, a^\dagger_{0A}\, a_{0A}$, with the impurity located in the central cell on the A sub-lattice and $U_0$ being the strength of the impurity potential (vacancy corresponds to $U_0 \gg /|t|$). The key quantity to compute is the full Green's function(GF) matrix $G^{Full}(\omega)$, which is related to the unperturbed Green's function $G^0(\omega)$ through the

Dyson's equation $G^{Full}(\omega) = G^0(\omega) + G^0(\omega) V G^{Full}(\omega)$. With the localized impurity potential, the Dyson's equation becomes

$$G^{Full}_{i,\alpha;j,\beta}(\omega) = G^0_{i,\alpha;j,\beta}(\omega) + U_0 G^0_{i,\alpha;0,A}(\omega) G^{Full}_{0,A;j,\beta}(\omega)$$

which upon inversion yields

$$G^{Full}_{i,\alpha;j,\beta}(\omega) = G^0_{i,\alpha;j,\beta}(\omega) + \{U_0 G^0_{i,\alpha;0,A}(\omega) G^0_{0,A;j,\alpha}(\omega)/(1 - U_0 G^0_{0,A;0,A}(\omega))\} \quad (6)$$

with $(i,\alpha)$ (and $(j,\beta)$) being the cell and sub-lattice indices. Here the propagator $G^0_{i,\alpha;j,\beta}(\omega)$ is given by

$$G^0_{i,\alpha;j,\beta}(\omega) = \Omega_{BZ}^{-1} \int d\mathbf{k} \, \exp[i\mathbf{k}\cdot(\mathbf{r}_{j,\beta} - \mathbf{r}_{i,\alpha})] G^0_{\alpha\beta}(\mathbf{k},\omega), \quad G^0_{\alpha\beta}(\mathbf{k},\omega) = \langle \mathbf{k}\alpha | G^0(\omega) | \mathbf{k}\beta \rangle.$$

In fact, we have used the Bloch basis function $|j\beta\rangle = N^{-1/2} \sum_k \exp(i\mathbf{k}\cdot \mathbf{r}_{j,\beta}) |\mathbf{k}\beta\rangle$ to arrive at the result above, where $\mathbf{r}_{i\alpha} = \mathbf{R}_i + \mathbf{r}_\alpha$ are the atom positions and N is the number of unit cells in the crystal. It follows from above that the full Green's function $G^{Full}_{0,A;0,A}(\omega)$ is given by

$$G^{Full}_{0,A;0,A}(\omega) = G^0_{0,A;0,A}(\omega) + \{U_0 G^0_{0,A;0,A}(\omega) G^0_{0,A;0,A}(\omega)/(1 - U_0 G^0_{0,A;0,A}(\omega))\}$$

$$= (F_0(\omega) - i\pi\rho_0(\omega)) + \{U_0 (F_0(\omega) - i\pi\rho_0(\omega))^2/(1 - U_0 F_0(\omega) + i\pi U_0 \rho_0(\omega))\}$$

$$= (F_0(\omega) - i\pi\rho_0(\omega)) + \{U_0(F_0(\omega)^2 - \pi^2\rho_0(\omega)^2 - 2i\pi\rho_0(\omega) F_0(\omega))/(1 - U_0 F_0(\omega) + i\pi U_0 \rho_0(\omega))\}. \quad (7)$$

All quantities of interest may be expressed in terms of the full Green's function $G^{Full}$, e. g., the local density of states (LDOS) at a specific site 'i' (with sub-lattice index 'α') is expressed as $\rho_{i\alpha}(\omega) = -\pi^{-1} \text{Im} G^{Full}_{i\alpha,i\alpha}(\omega)$. From Eq. (7), the LDOS at the impurity site has an especially simple form

$$\rho_{0A}(\omega) = \rho^{(0)}_{0A}(\omega) / [(1 - U_0 F_0(\omega))^2 + (\pi \rho^{(0)}_{0A}(\omega) U_0)^2] \quad (8)$$

where $\rho^{(0)}_{0A}(\omega) = -\pi^{-1} \text{Im} G^0_{i\alpha,i\alpha}(\omega)$. Now in view of Eqs. (7) and (8), the contribution of the impurity has a sharp peak at the resonance energy $E_0$ satisfying the resonance condition $(1 - U_0 F_0(E_0)) = 0$. Near the resonance we Taylor expand $F_0(\omega)$ as $F_0(\omega) = U_0^{-1} + F'_0(E_0)(\omega - E_0) + (1/2) F''_0(E_0)(\omega - E_0)^2 + \ldots$. This enables us to write

$$|t|\rho_{0A}(\omega) = |t|\rho^{(0)}_{0A}(\omega) (|t|F'_0(E_0))^{-2}$$

$$\times [((\omega/|t|) - (E_0/|t|))^2 + (1/4)(|t|^2 F''_0(E_0)/(|t|F'_0(E_0)))^2 ((\omega/|t|) - (E_0/|t|))^4$$

$$+ (|t|^2 F''_0(E_0) / |t| F'_0(E_0)) ((\omega/|t|) - (E_0/|t|))^3 + (\pi |t| \rho^{(0)}_{0A}(\omega) (U_0/|t|))^2 ]^{-1}. \quad (9)$$

The entire exercise above allows us to calculate the full propagator $G^{Full}_{0,A;\,0,A}(\omega)$ from (4). The plot of $[|t| \rho_{0A}(\varepsilon/|t|)]$ as a function of $(\varepsilon/|t|)$ is shown in Figure 2. We have assumed in Figure 2(a) the following numerical values: The level-broadening factor $(\acute{\Gamma}/|t|) = 0.05$, the ratio $(t'/|t|) = 0.08$, $E_0/|t|$ (blue curve) = 0.5, $E_0/|t|$ (green curve) = 0.1, $U_0/|t|$ (blue curve) = 1, and $U_0/|t|$ (green curve) = 10. The peaks correspond to broad resonances. The particle-hole symmetry is approximately present. Our result implies that introducing a vacancy in an otherwise perfect lattice, immediately creates a zero energy mode state (ZMS). This is important result because ZMS's are created precisely at the Fermi level, and have this peculiar topological localization determining that they should live in just one of the lattices (see next section). In Figure 2(b) we have shown the plot of $[|t| \rho_{0A}(\varepsilon/|t|)]$ as a function of $(\varepsilon/|t|)$ for $(t'/|t|) = 0.2$. The rest of the numerical values are the same as in Figure 2(a). The suppression of ZMS occurs due to increase in the value of $(t'/|t|)$.

### 4. Wave function localization and moment formation

As regards the wave function of the impurity state, it is seen above that if 2NN interaction present in a bipartite lattice is much weaker than the NN interaction, a single vacancy in one sub-lattice (which may be referred to as the minority sub-lattice now since there is one atom less) introduces a state at the zero of energy (**ZMS**) and the result basically follows from the approximate particle-hole symmetry. We shall now see that this state does not live on the minority sub-lattice. This result follows by examining the perturbed wave function in the presence of the impurity potential following the Lippmann-Schwinger(LS) equation.

In view of the LS equation given in section 1 and the fact that $G^0_{0,A;\,0,A}(\omega) = \Omega_{BZ}^{-1} \int d\mathbf{k} \exp[i\mathbf{k}.(\mathbf{r}_{0,A} - \mathbf{r}_{0,A})] G^0_{AA}(\mathbf{k},\omega)$ is given by $[F_0(\omega) - i\pi \rho^{(0)}_{0A}(\omega)]$, we find that the perturbed wave function at a site i and the sub-lattice α is given by

$$\psi_{i\alpha} = \psi^0_{i\alpha} + \{ U_0 G^0_{i\alpha,0A}(\omega) \psi^0_{0A} / (1 - U_0 G^0_{0A,0A}(\omega)) \}$$

$$= \psi^0_{i\alpha} + [(U_0 G^0_{i\alpha,0A}(\omega) \psi^0_{0A}) / \{(1 - F_0 U_0) + i\pi \rho^{(0)}_{0A}\}].$$

Under the resonance condition this yields

$$\psi_{i\alpha} = \psi^0_{i\alpha} + \{G^0_{i\alpha,0A}(\omega) \psi^0_{0A} / (i\pi \rho^{(0)}_{0A}(\omega))\}.$$

From this equation one obtains

$$\psi_{iA} = \psi^0_{iA} + \{G^0_{iA,0A}(\omega)\psi^0_{0A}/(i\pi\rho^{(0)}_{0A}(\omega))\}$$

$$= \psi^0_{0A}\{F_0(\omega)/(i\pi\rho^{(0)}_{0A}(\omega))\} = \psi^0_{0A}\{(U_0/|t|)^{-1}/(i\pi|t|\rho^{(0)}_{0A}(\omega))\}$$

where the last line follows from the fact that for ZMS $F_0(\omega) = U_0^{-1}$. Since the strong scalar potential $((U_0/|t|) \gg 1)$ is induced by the vacancy it follows from above that ZMS cannot live on the minority sub-lattice. As regards the issue of wave function residing on the majority sub-lattice, one may refer to a seminal paper by Castro Neto et al. [4] where the authors have proved that, if there is an imbalance in the number of atoms in the two sub-lattices in a bipartite lattice, viz., $n = N_B - N_A > 0$, there are n number of degenerate solutions with the eigenvalue equal to the on-site energy of the majority sub-lattice with the wave functions residing entirely on this sub-lattice ( the wave function decays inversely with distance[5]). The proof is based on the rank-nullity theorem in linear algebra; we shall not repeat this proof here. We, however, note that the theorem has an important implication: If the more than one vacancies are present one in each cell of the system connected by the lattice translational vectors, i.e. all are located on the same sub-lattice forming the minority sub-lattice, and if 'n' is the number of unit cells in the crystal, then according to the theorem, there should be 'n' zero-modes in the Brillouin zone (BZ) which is also precisely the number of Bloch momentum points in the BZ. These states then form a dispersion-less band in the tight-binding calculations at zero energy.

We now wish to provide an instructive argument to clarify how the magnetic moment formation takes place with an isolated vacancy. We first recall that the atomic configuration of the carbon atoms is $1s^2 2s^2 2p^2$. However, in graphene, the electronic configuration is $1s^2 2s 2p^3$. In Figure 3 we have shown the $sp^2$-hybridized orbitals of a carbon atom in graphene. Due to the $sp^2$-hybridization the atoms form a hexagonal lattice with unit cell consisting of two atoms with σ - bond on the two sub-lattices A and B. The $p_z$-orbitals with π-bond do not participate in the $sp^2$-hybridization. The electrical conduction is basically due to the hopping between the $p_z$ orbitals. In fact, graphene has only one conduction electron per atom, which is in the $2p_z$ state. Now without a vacancy, the $sp^2$-states are lodged away from the Fermi energy ($E_F$) due to strong interaction with neighboring orbitals along the C-C bonds. However, when a vacancy is present, the three $sp^2$-orbitals of each of the three nearest-neighbor carbon atoms, forming a carbon triangle surrounding the vacancy, have their bonding partners missing, so that they occur near $E_F$, with their on-site energies below the π orbital energies because of the s-orbital component present in the σ-states. The on-site Coulomb interaction would prevent the occupation of a fourth $sp^2$- state, for that would cause a double occupancy of a dangling bond on the carbon triangle involving high energy cost. Thus, the remaining electron occupies the π states. The Hund's coupling with the σ electrons will spin polarize the π state, with the energy of the majority-spin state now being lodged below $E_F$. This leads to a magnetic moment (~$\mu_B$) formation. In conclusion, we wish to note that since the system at the Fermi level has low electronic density, there is poor screening of such magnetic moments. This may lead to a high Curie temperature for such vacancy-induced moments.

**References**

**1.** A. H. Castro Neto, F. Guinea, N. M. R. Peres, K. S. Novoselov, and A. K. Geim, Rev. Mod. Phys. 81, 109 (2009). **2.** M. I. Katsnelson, K. S. Novoselov, and A. K. Geim, Nature Phys. 2, 620-625 (2006). **3.** M. I. Katsnelson, Eur. Phys. J. B 51, 157-160 (2006). **4.** V. M. Pereira, J. M. B. Lopes dos Santos, and A. H. Castro Neto, Phys. Rev. B 77, 115109 (2008). **5.** V. M. Pereira, F. Guinea, J. M. B. L. dos Santos, N. M. R. Peres, and A. H. Castro Neto, Phys. Rev. Lett. 96, 036801(2006). **6.** M. Sherafati and S. Satpathy,arXiv:1103.4679. **7.** M. M. Ugeda, I. Brihuega, F. Guinea, and J. M. Gomez-Rodriguez, Phys. Rev. Lett. 104, 096804(2010).

**Figures and Captions**

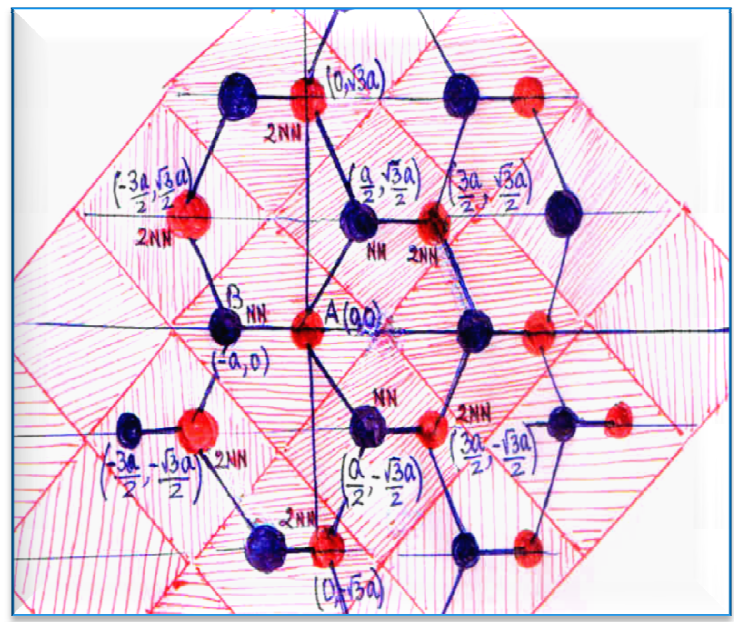

**Figure A.** A diagrammatic representation of the 2-D honey-comb (mono-layer) graphene lattice structure. The lattice is bipartite and composed of two triangular sub-lattices, commonly labeled A and B, so that there are two atoms per unit cell. The vertices of this type of structure do not form a Bravais lattice. The array of points do not have the same appearance when viewed from A and B; the view from B appear to be rotated. The distance between nearest neighbor atoms is a ≈ 1.42°A. The red and blue circles, respectively, indicate carbon atoms on the A and B sub-lattices. The hatched regions (rhombus) indicate unit cells.

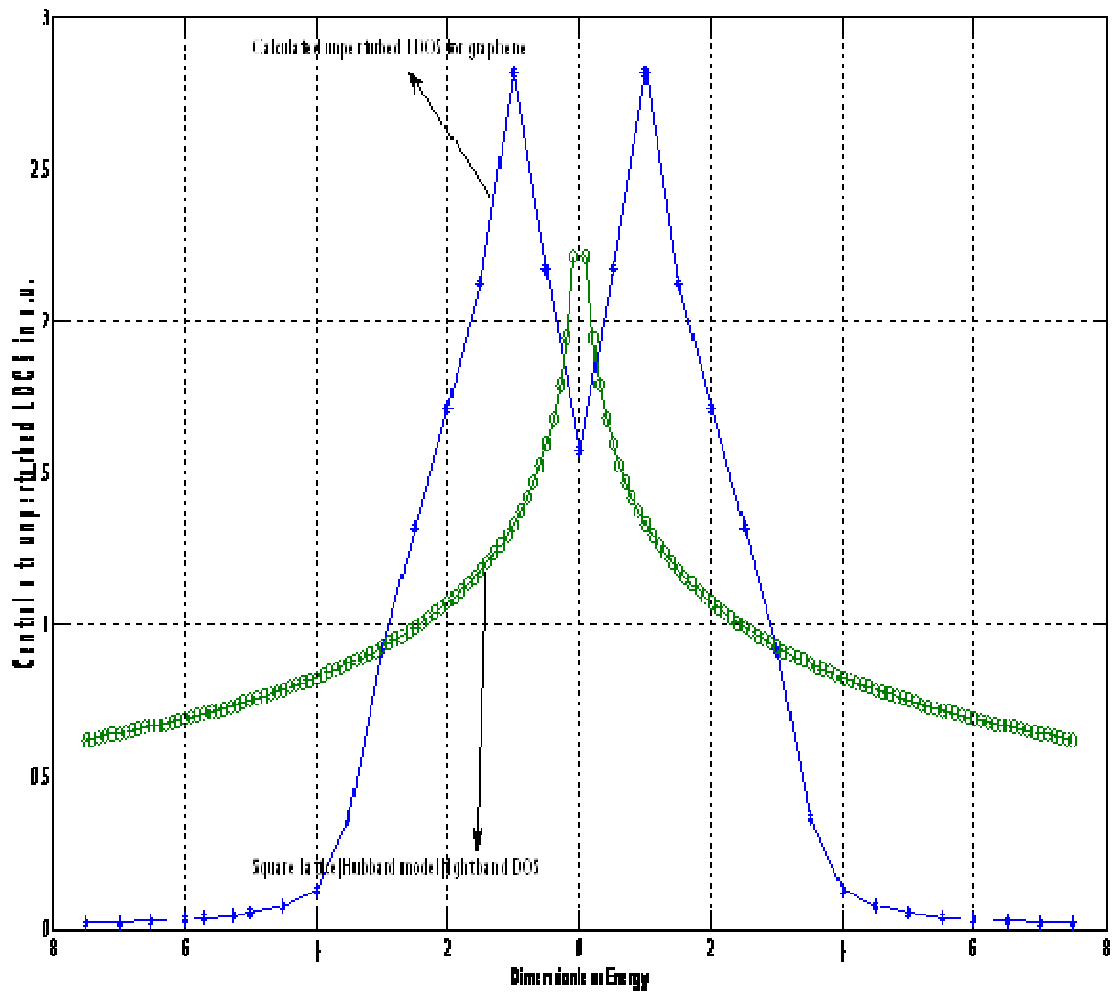

(a)

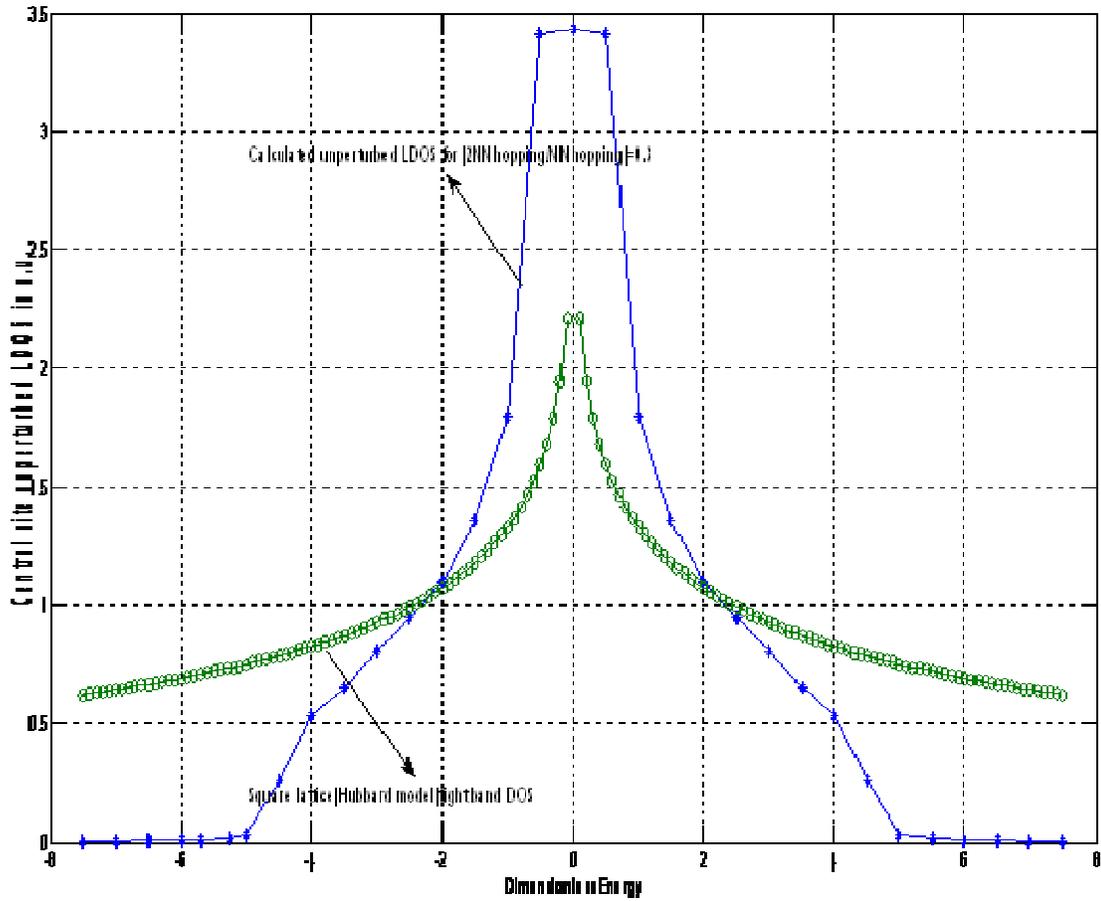

(b)

**Figure 1.** (a) The plot (blue curve) of $[\rho^{(0)}_{0A}(\varepsilon/\ )]$ as a function of $(\varepsilon/\ )$. The level-broadening factor $(\acute{\Gamma}\ )=0.2$. The ratio $(t'/\ )=0.08$. The green curve is the square lattice tight band DOS in the Hubbard model. (b) The plot (blue curve) of $[\rho^{(0)}_{0A}(\varepsilon/\ )]$ as a function of $(\varepsilon/\ )$ for the level-broadening factor $(\acute{\Gamma}\ )=0.05$ and the ratio $(t'/\ )=0.3$. The plot is qualitatively almost similar to the green curve corresponding to the square lattice tight band DOS in the Hubbard model.

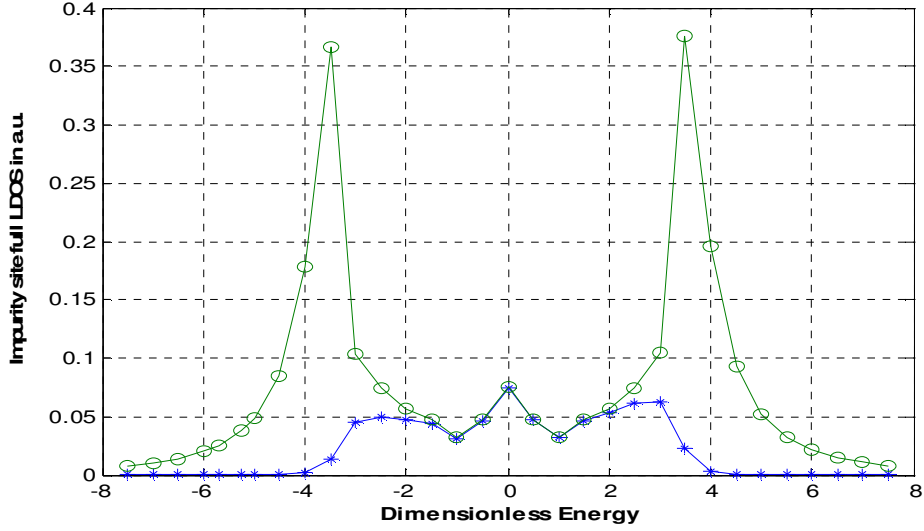

(a)

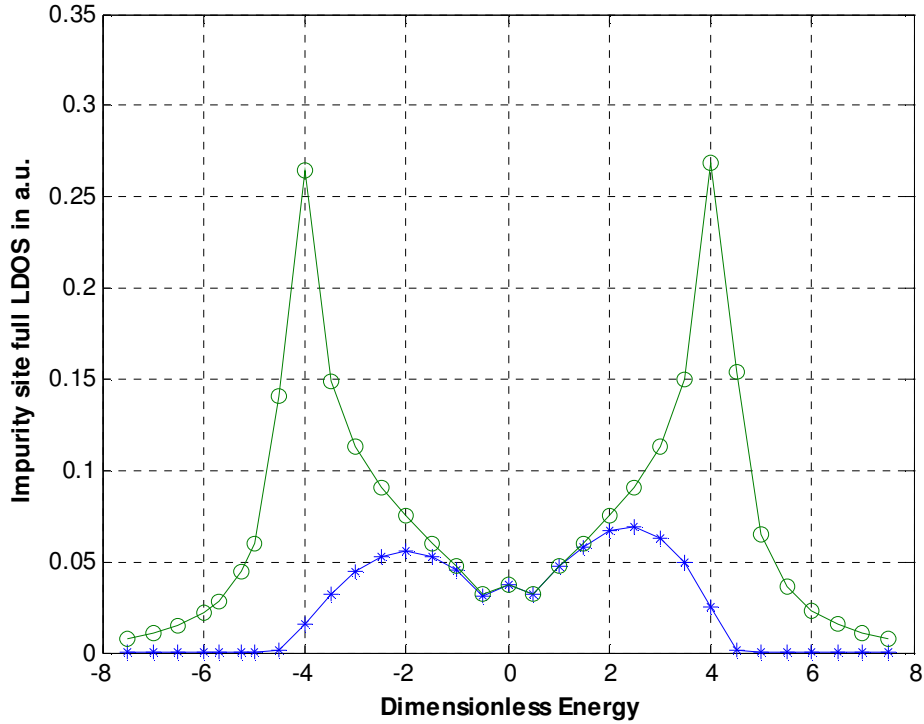

(b)

**Figure 2.** (a)The plot of $[|t|\rho_{0A}(\varepsilon/|t|)]$ as a function of $(\varepsilon/|t|)$. The ratio $(t'/|t|) = 0.08$, $E_0/|t|$ (blue curve) = 0.5, $E_0/|t|$ (green curve) = 0.1, $U_0/|t|$ (blue curve) = 1, and $U_0/|t|$ (green curve) = 10. The peaks correspond to broad resonances. The particle-hole symmetry is approximately present. As the impurity potential $U_0$ is increased to infinity, the solution $E_0$ approaches the zero of energy resulting in the so-called "zero-mode" state.

(b) The plot of [ ρ₀A(ε/ )] as a function of (ε/ ). The ratio (t'/ = 0.2 and rest of the numerical values are the same as in (a), The particle-hole asymmetry and the suppression of the ZMS are the important features.

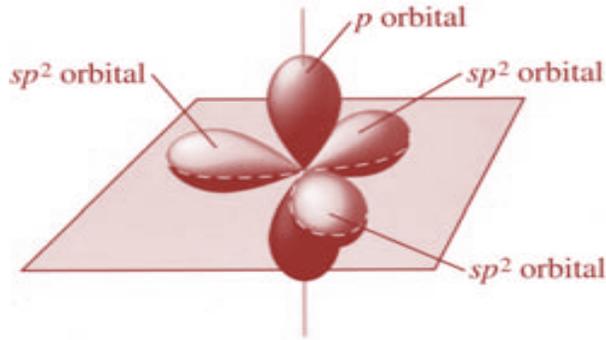

**Figure3.** The sp² –hybridized orbitals of a carbon atom in graphene(s is like a ball, p is like a dumb-bell, and in sp hybrids one electron is waiting for another to fill in(bonding)). There is one electron in each of the three sp² –hybridized orbitals (say, in x-y plane; the angle between any two such orbitals is 120°) plus one electron in $p_z$ orbital(perpendicular to the x-y plane). Very strong covalent bonds are formed through sp²-orbitals when neighboring carbon atoms come close.